\documentclass[11pt]{article}
\newcommand{\be}{\begin{equation}}
\newcommand{\ee}{\end{equation}}
\newcommand{\bea}{\begin{eqnarray}}
\newcommand{\eea}{\end{eqnarray}}
\newcommand{\bean}{\begin{eqnarray*}}
\newcommand{\eean}{\end{eqnarray*}}
\newcommand{\brray}{\begin{array}}
\newcommand{\erray}{\end{array}}
\newcommand{\ben}{\begin{equation}{nonumber}}
\newcommand{\een}{\end{equation}{nonumber}}

\newtheorem{dfn}{Definition}[section]
\newtheorem{thm}[dfn]{Theorem}
\newtheorem{lmma}[dfn]{Lemma}
\newtheorem{ppsn}[dfn]{Proposition}
\newtheorem{crlre}[dfn]{Corollary}
\newtheorem{xmpl}[dfn]{Example}
\newtheorem{rmrk}[dfn]{Remark}

\newcommand{\bdfn}{\begin{dfn}}
\newcommand{\bthm}{\begin{thm}}
\newcommand{\blmma}{\begin{lmma}}
\newcommand{\bppsn}{\begin{ppsn}}
\newcommand{\bcrlre}{\begin{crlre}}
\newcommand{\bxmpl}{\begin{xmpl}}
\newcommand{\brmrk}{\begin{rmrk}}

\newcommand{\edfn}{\end{dfn}}
\newcommand{\ethm}{\end{thm}}
\newcommand{\elmma}{\end{lmma}}
\newcommand{\eppsn}{\end{ppsn}}
\newcommand{\ecrlre}{\end{crlre}}
\newcommand{\exmpl}{\end{xmpl}}
\newcommand{\ermrk}{\end{rmrk}}

\newcommand{\al}{\alpha}

\newcommand{\gma}{\gamma}

\newcommand{\cla}{{\cal A}}

\newcommand{\cld}{{\cal D}}

\newcommand{\clh}{{\cal H}}

\def\a*{{\cal A}_{h,*}}
\def\B{{\cal B}(h)}
\def\B1{{\cal B}_1(h)}
\def\b{{\cal B}^{s. a. }(h)}
\def\b1{{\cal B}^{s. a. }_1(h)}

\newcommand{\ot}{\otimes}

\newcommand{\raro}{\rightarrow}

\begin{document}


\begin{center}
{\bf{\large Some Noncommutative Geometric Aspects of $SU_q(2)$}}\\
by\\  Debashish Goswami \\ \emph{Indian Statistical Institute,
Kolkata Centre; 203, B.T. Road, Kolkata-700108, India.\\
 email
: goswamid@hotmail.com}
\end{center}

\begin{abstract}
We study various noncommutative geometric aspects  of the compact quantum
 group  $SU_q(2)$ for positive $q$ (not equal to $1$),
 following the suggestion of Connes and his coauthors (\cite{CL},
  \cite{CD}) for considering  the so-called true Dirac operator.
   However, it turns out that the method of the above references do not extend
    to the case of positive (not equal to $1$) values of $q$ in the sense that
     the true Dirac operator does not have bounded commutators with ``smooth" algebra
      elements in this case, in contrast to what happens for complex $q$ of modulus $1$.
       Nevertheless,  we
   show how to obtain the canonical volume form, i.e. the Haar state,
using
     the true Dirac operator.   

 \end{abstract}

Sub. class : 81R50, 58B34, 81R60\\

\section{Introduction}
Noncommutative geometry (introduced by A. Connes (\cite{C1}, \cite{C2}))
 and the theory of quantum groups (introduced and studied by Drinfeld
 \cite{D}, Jimbo \cite{J},
 Woronowicz \cite{W1} and others) are two important and rapidly growing areas
 of the so-called ``noncommutative mathematics", both having close interactions with
 and applications to diverse branches of mathematics and mathematical physics
 (e.g. quantum gravity).
 Since almost all the popular and well-known examples of quantum groups are in some sense
 ``deformed" or ``twisted" versions of some classical Lie groups, which
  are also differentiable manifolds and  play a central
 role in classical differential geometry, it is quite natural to investigate the relation
 between Connes' noncommutative geometry and the theory of quantum groups. Such a connection
 was made by Woronowicz himself in his pioneering paper \cite{W1}, where he formulated and studied
 the notions of differential forms and de-Rham cohomology in the context of $SU_q(2)$.
 Since then, many authors including Woronowicz, Podles and others have studied such
 questions quite extensively and a rich theory of covariant (and bicovariant) differential
 calculus has emerged. However, a number of questions in the interface of noncommutative geometry
 and quantum groups still remains open.
 In particular, the ``naive" approach (in the language of \cite{CL}) taken by some authors including Bibikov, Kulish,
 Woronowicz, Podles and others to build a suitable theory of noncommutative differential
 geometry in the context of quantum groups suffered from some serious disadvantages.
  One of the major difficulty is that various ``naive" Dirac operators introduced by these authors
 (e.g. \cite{BK}) do not have the property that  the commutators of such Dirac operators
 with the elements of a distinguished dense subalgebra are bounded. Without this perperty of having bounded
 commutators, it is not possible to carry out most of the canonical general constructions
 done in Connes' framework. Furthermore, the spectral properties of such Dirac operators or the associated
 `` naive" Laplacians are quite strange in some sense.

To overcome the above difficulty, Connes and his coauthors (see \cite{CL}, \cite{CD})
 have recently suggested a better candidate of Dirac operator, termed as the ``true"
 Dirac operator in \cite{CL}, in the context of some quantum groups, including $SU_q(2)$.
 However, their framework includes only the case where $q$ is complex of modulus $1$,
 but we are interested in the case where $q$ is positive and not equal to $1$.
  We shall show that a similar definition of Dirac operator does {\bf not} work
 for such $q$'s. In fact, if we denote by $D$ the true Dirac operator suggested by \cite{CL}
  and \cite{CD} in this case, then $[D,a]$ is not bounded for $a$ in the distinguished
   ``smooth" algebra, even though $[|D|,a]$ is bounded. Nevertheless, we show how to obtain
    the Haar state from the Laplacian, i.e. $D^2$.

\section{Notation and Preliminaries}
 Before we enter into the discussion on $SU_q(2)$, let us discuss little bit about
 classcical compact Lie groups. Let $G$ be a compact $d_0$-dimensional Lie
 group, equipped with an invariant Riemannian metric, and $h$ be the Hilbert
 space $L^2(G,\mu)$, where $\mu$ is the normalized Haar measure. If we consider
 $h$ as canonically embedded in the Hilbert space of differential forms, and if $C$
 denotes the restriction of the Hodge Laplacian onto $h$, then $Lim_{t \raro 0+}
 t^{d_0/2} Tr(exp(-tC))$ exists and is nonzero, and furthermore, $Tr(fexp(-tC))=
 v(f) Tr(exp(-tC))$ for any continuous function $f$ on $G$, viewed as a multiplication operator
 on $h$, and where $v(f)=\int f d \mu$. However, as we shall shortly see, such classical intuition
  is  no longer valid for quantum groups, but we can recover some aspects of it  by introducing an appropriate
  modified formulation of noncommutative  geometry.

 Let us now describe the compact quantum groups $SU_q(2).$
Let $\cla$ be the $C^*$-algebra associated with the quantum group $SU_q(2)$ as defined
 in \cite{W1} ($q$ positive number),   i.e. the universal unital $C^*$-algebra
 generated by $\al, \gma, \al^*, \gma^*$ satisfying $\al^* \al+\gma^* \gma= 1,
 \al \al^* +q^2 \gma^* \gma=1, \gma^* \gma=\gma \gma^*, \al \gma=q\gma \al,
 \al \gma^*=q \gma^* \al.$ We shall take $q$ to be
 greater than $1$, but remark that  all our results will be valid also when $0<q<1$,
 which can be seen by obvious modifications at a few steps of the proofs of some of the
  results. Let $\cla^\infty$ denote
  the $\ast$-algebraic span of $\al, \gma$. Let $\Phi, \kappa$ and $\epsilon$ denote
 the coproduct, antipode and counit respectively, as defined in \cite{W1}.
 We recall the construction of the
 normalized Haar state $\psi$ on $\cla$, and let $h$ be the $L^2$ space associated
 with this state.
 As usual, we denote by $t^n_{ij}, i,j, =-n,...n; n=0, \frac{1}{2},1,\frac{3}{2},...$
 the matrix elements of the unitary irreducible co-representations, so that
 $span \{ t^n_{ij},(t_{ij}^n)^*: n,i,j \}=\cla^\infty$, and $\Phi(t^n_{ij})=\sum_k t^n_{ik} \ot
 t^n_{kj}$, $\kappa(t^n_{ij})= (t^n_{ji})^*$. To avoid any confusion, we choose the following
 explicit definition of these matrix elements. Let $x_{nj}=\al^{n+j} (\gma^*)
^{n-j}$, $n=0, \frac{1}{2}, 1,...; j=-n,...,n$, and let $y_{nj}=
 \frac{x_{nj}}{\psi(x_{nj}^*x_{nj})^{\frac{1}{2}}}.$ Then we define $t^n_{ij}$
  by the relation $\Phi(y_{nj})=\sum_i y_{ni} \otimes t^n_{ij}.$
   It is easy to verify that $<t^m_{ij}, t^n_{kl}> :=
 \psi((t^m_{ij})^*t^n_{kl})=\delta_{ik} \delta_{jl} \delta_{mn} [2n+1]^{-1}_q
  q^{2i}$, where $\delta_{rs}$ is  the Kronecker delta symbol and $[r]_q:=
  \frac{q^r-q^{-r}}{q-q^{-1}}$.
     Furthermore, $\psi(t^m_{ij}(t^n_{kl})^*)=\delta_{ik}\delta_{jl}
 \delta_{mn}[2n+1]^{-1}_q q^{-2j}.$ We consider the orthonormal basis of $h$
 given by $\{ \tilde{t^n_{kl}}:=t^n_{kl} [2n+1]^{\frac{1}{2}}_q q^{-k} \}$.

\section{The ``true" Dirac operator on $SU_q(2)$ and some geometric
consequences }
 We shall follow the suggestion of \cite{CL} to define and
study
 the ``true" Dirac
 operator and formulate an appropriate modification of Connes' spectral triple.
 First of all, let us recall the construction of the ``naive" Dirac operator
 as in \cite{BK}, to be denoted by $Q$.
Let $\clh$ denote the Hilbert space $C^2 \ot h$, $h=L^2(\cla, \psi)$ as defined
 earlier. We denote by $\pi_l, l=0, \frac{1}{2}, 1, \frac{3}{2},...$ the
 $(2l+1)$-dimensional irreducible corepresentation of $SU_q(2)$ respectively,
 and identify the first tensor component $C^2$ of $\clh$ with the space of
 $\pi_{\frac{1}{2}}$, denoting by $e_+\equiv |\frac{1}{2},\frac{1}{2}>,
 e_{-} \equiv |\frac{1}{2},-\frac{1}{2}>$
 the canonical orthonormal basis associated with this corepresentation.
  We define the following distinguished vectors of $\clh$ :
$$ v^{l,+}_{ij}:= C^{\frac{1}{2},l,l+\frac{1}{2}}_{\frac{1}{2},j-\frac{1}{2},
 j} e_+ \ot \tilde{t}^l_{i,j-\frac{1}{2}}+
 C^{\frac{1}{2},l,l+\frac{1}{2}}_{-\frac{1}{2},j+\frac{1}{2},
 j} e_- \ot \tilde{t}^l_{i,j+\frac{1}{2}},$$
$$v^{l,-}_{ij}:= C^{\frac{1}{2},l,l-\frac{1}{2}}_{\frac{1}{2},j-\frac{1}{2},
 j} e_+ \ot \tilde{t}^l_{i,j-\frac{1}{2}}+
 C^{\frac{1}{2},l,l-\frac{1}{2}}_{-\frac{1}{2},j+\frac{1}{2},
 j} e_- \ot \tilde{t}^l_{i,j+\frac{1}{2}};$$
where $C^{a,b,c}_{k,l,m}$ denote the $q$-Clebsch-Gordan coefficients,
 mentioned in \cite{BK} (the notation of \cite{BK} was
 $\left[ \begin{array}{cc} a,b,c\\ k,l,m \end{array} \right]_q$ for
 our $C^{a,b,c}_{k,l,m}$).  The ``naive" Dirac operator $Q$ (denoted by
 \cite{BK} as $D_q$) is a self-adjoint operator having $C^2 \ot_{alg} \cla^\infty$
 in its domain and having $v^{l,+}_{ij}, v^{l,-}_{ij}, i,j=-l,-l+1,...,l; l=0, \frac{1}{2},...
$ as a complete set of eigenvectors with eigenvalues $[l]_{q^2}$ and $-[l+1]_{q^2}$
 corresponding to $v^{l,+}_{ij}$ and $v^{l,-}_{ij}$ respectively.
 We now define the ``true" Dirac operator to be the self-adjoint operator
 $D$ with the same eigenvectors as above, but the corresponding eigenvalues
 being $l+\frac{1}{2}$ and $-(l+\frac{1}{2})$ respectively. It is to be noted that
 $D$ is related to $Q$ by the relation $[D-\frac{1}{2}I]_{q^2}=Q$, and thus
 this $D$ is a trivial modification of the one suggested by \cite{CL}, by just adding
 $\frac{1}{2}I$ to their original prescription. We let $\cla$ act on $\clh$
 by $a \mapsto (I_2 \ot a)$, and we also denote $(I_2 \ot a)$ by $a$, without
 any possibility of confusion. Our first task is to verify whether
  the remark made by \cite{CL}
  holds true in the present context, i.e. for $q$ positive (not $1$).
\bthm
\label{dbdd}
$[|D|,a]$ is bounded for any $a \in \cla^\infty$.\\

\ethm
{\it Proof :-}\\
(1) Clearly, $|D|$ is of the form $I_2 \ot A$, where $A$ is the operator with
 $\tilde{t^n_{ij}}$'s as a complete set of eigenvectors with $n+\frac{1}{2}$ as
 the corresponding eigenvalue. Thus, we have to show  $[A,a]$ is bounded for
 $a \in \cla^\infty$. It is enough to show it for $a=t^n_{ij}$, or equivalently,
 for $\tilde{t^n_{ij}}$, which is a constant multiple of $t^n_{ij}$. Let us
 fix some $n=n_0,i=i_0,j=j_0$. We note that $$\tilde{t^n_{ij}} \tilde{t^m_{kl}}
=\sum_{p=|n-m|,|n-m|+1,...n+m} B(n,m,p;i,j,k,l)
\tilde{t}^p_{i+k,j+l}. $$ Let $v \in h$ be given by
$v=\sum_{n,i,j} v^n_{ij} \tilde{t^n_{ij}}$.
 Let us make the following notational  convention : $\tilde{t^n_{ij}}$ will
 be set equal to $0$ if any of the indices $i,j$ falls outside the
 range $\{ -n,-n+1,...n \}$. We then have,
\bean \lefteqn{ \| [A,\tilde{t^{n_0}_{i_0j_0}}]v\|^2}\\ &=& \|
\sum_{n,i,j}   v^n_{ij} \sum_{p=|n-n_0|,|n-n_0|+1,...n+n_0}
B(n_0,n,p;i_0,j_0,i,j)(p-n) \tilde{t^p}_{i_0+i,j_0+j} \|^2\\ &=&
\sum_{p,i,j}| \sum_{ n
: |n-n_0| \leq p \leq n+n_0}
 v^n_{ij} B(n_0,n,p;i_0,j_0,i,j) (p-n)|^2.\\
\eean
Clearly, the number of $n$ satisfying $|n-n_0| \leq p \leq n+n_0$ is at
 most $2n_0+1$, and furthermore, for any such $n$, $|p-n| \leq n_0$, so that
 \bean
 \lefteqn{ | \sum_{  n : |n-n_0| \leq p \leq n+n_0}
 v^n_{ij} B(n_0,n,p;i_0,j_0,i,j) (p-n)|^2}\\
 & \leq& (2n_0+1)n_0^2
\sum_{  n : |n-n_0| \leq p \leq n+n_0} |v^n_{ij}|^2
 (B(n_0,n,p;i_0,j_0,i,j))^2 .
 \eean
  From this, we obtain,
\bean
\lefteqn{\| [A,\tilde{t^{n_0}_{i_0j_0}}]v\|^2}\\
&\leq& (2n_0+1)n_0^2 \sum_{p,i,j} \sum_{n : |n-n_0| \leq p \leq n+n_0} |v^n_{ij}|^2
 (B(n_0,n,p;i_0,j_0,i,j))^2 \\
&=& (2n_0+1)n_0^2 \sum_{n,i,j} |v^n_{ij}|^2 \sum_{p : |n-n_0| \leq p \leq n+n_0}
 (B(n_0,n,p;i_0,j_0,i,j))^2.\\
\eean
Now, we note that for any fixed $n,i,j$, $\sum_{p : |n-n_0| \leq p \leq n+n_0}
 (B(n_0,n,p;i_0,j_0,i,j))^2=\| \tilde{t^{n_0}_{i_0j_0}} \tilde{t^n_{ij}}\|^2
 \leq C^2 \|\tilde{t^n}_{ij} \|^2 =C$, where $C=\| \tilde{t^{n_0}_{i_0j_0}}\|$ is a constant.
 It is clear that $\| [A,\tilde{t^{n_0}_{i_0j_0}}]v\|^2 \leq C^2 \|v\|^2,$ which
 completes the proof of the theorem. 
\fbox \\ 

\brmrk
 However, it is not true that $[D,a]$ is bounded for $a$ as above; in particular 
 for $a=\tilde{t}^{\frac{1}{2}}_{\frac{1}{2},\frac{1}{2}}$, $[D,a]$ is not bounded. 
 It is easy to see
 that for $a=\tilde{t^{\frac{1}{2}}_{\frac{1}{2},\frac{1}{2}}}$ and
 any $l \geq \frac{1}{2}$,
 $$a v^{l,+}_{ij}= \sum_{m} b_{m}^+(i,j) v^{m,+}_{i+\frac{1}{2},j+\frac{1}{2}}
+\sum_{m}  b_{m}^-(i,j) v^{m,-}_{i+\frac{1}{2},j+\frac{1}{2}},$$
 where the range of $m$ is    $\{ l-\frac{1}{2},l+\frac{1}{2} \}$,
   and $$b_{m}^\epsilon(i,j)
 = \sum_{m_1=\frac{1}{2},-\frac{1}{2}}
  C^{\frac{1}{2},l,l+\frac{1}{2}}_{m_1,j-m_1,j} C^{\frac{1}{2},l,m}_{\frac{1}{2},
   i, i+\frac{1}{2}}C^{\frac{1}{2},l,m}_{\frac{1}{2},
j-m_1,j-m_1+\frac{1}{2}}
 C^{\frac{1}{2},m,m+\frac{1}{2}\epsilon}_{m_1,j+\frac{1}{2}-m_1,j+\frac{1}{2}} \frac{[2]_q^
 {\frac{1}{2}}[2l+1]_q^{\frac{1}{2}}}{[2m+1]_q^{\frac{1}{2}}},$$
 for $\epsilon =+,-$. Now, it is straightforward to verify that
 $[D,a]v^{l,+}_{ij}=\sum_{m} \{ (m-l)b^+_{m}(i,j) v^{m,+}_{i+\frac{1}{2},j+
\frac{1}{2}}-
 (m+l) b^-_{m}(i,j) v^{m,-}_{i+\frac{1}{2},j+\frac{1}{2}} \}$.
 Since
  $ v^{m,\epsilon}_{i+\frac{1}{2},j+\frac{1}{2}}$ are clearly orthogonal for different
 choices of the pair  $(m,\epsilon)$, it follows that
     $\| [D,a]v^{l,+}_{ij}\|^2 =\sum_{m, \epsilon} |m-\epsilon l|^2
 |b^{\epsilon}_{m}(i,j)|^2 \|v^{m,\epsilon}_{i+\frac{1}{2},j+\frac{1}{2}}\|^2.$
  Hence, if $[D,a]$ is bounded, the quantity \\ $\frac{|l+\epsilon^\prime
 \frac{1}{2}-\epsilon l|^2|b^{\epsilon}_{l+\epsilon^\prime
 \frac{1}{2}}(i,j)|^2
 \|v^{l+\epsilon^\prime\frac{1}{2},\epsilon}_{i+\frac{1}{2},j+\frac{1}{2}}\|^2}
{\| v^{l,+}_{ij} \|^2}$ must be bounded as $l,i,j, \epsilon, \epsilon^\prime$
 ($\epsilon^\prime=+,-$) are allowed
 to vary. 

We note  the following values of the Clebsch-Gordan coefficients,
 which are taken from \cite{BL}, with the care that one has to replace $q$
 in the formulae of that book by $q^{-2}$ to apply it to our case.
 $$ C^{\frac{1}{2},l,l+\frac{1}{2}}_{\frac{1}{2},m,m+\frac{1}{2}}=
 q^{\frac{l-m}{2}} \frac{[l+m+1]_q^{\frac{1}{2}}}{[2l+1]_q^{\frac{1}{2}}},
 C^{\frac{1}{2},l,l+\frac{1}{2}}_{-\frac{1}{2},m,m-\frac{1}{2}}=
 q^{-\frac{l+m}{2}} \frac{[l-m+1]_q^{\frac{1}{2}}}{[2l+1]_q^{\frac{1}{2}}},$$
$$ C^{\frac{1}{2},l,l-\frac{1}{2}}_{\frac{1}{2},m,m+\frac{1}{2}}=
 q^{-\frac{l+m+1}{2}} \frac{[l-m]_q^{\frac{1}{2}}}{[2l+1]_q^{\frac{1}{2}}},
 C^{\frac{1}{2},l,l-\frac{1}{2}}_{-\frac{1}{2},m,m-\frac{1}{2}}=
 -q^{\frac{l-m+1}{2}} \frac{[l+m]_q^{\frac{1}{2}}}{[2l+1]_q^{\frac{1}{2}}}.
 $$
Using these expressions and after some straightforward simplification, we see that
 the coefficient $b^-_{l+\frac{1}{2}}(i,j)$ is given by,
$$  b^-_{l+\frac{1}{2}}(i,j)=\frac{q^{\frac{l-3j-\frac{1}{2}}{2}}
[l-j+\frac{1}{2}]_q^ {\frac{1}{2}}}{[2l+1]_q
[2l+2]_q^{\frac{1}{2}}} \left( [l+j+\frac{1}{2}]_q
 -[l+j+\frac{3}{2}]_q \right)C^{\frac{1}{2},l,l+\frac{1}{2}}_{\frac{1}{2},i,i+\frac{1}{2}}
  \frac{[2]_q^
 {\frac{1}{2}}[2l+1]_q^{\frac{1}{2}}}{[2l+2]_q^{\frac{1}{2}}}.$$
 Now, putting $j=-l-\frac{1}{2},i=l$ in the above expression, it is easy to see that
 the absolute value of the above quantity converges to a finite nonzero number  as $l \raro \infty$, and thus
 there is a strictly positive lower bound of this sequence for large $l$.
 It is also simple to check that $\frac{\| v^{l+\frac{1}{2},-}_{i+\frac{1}{2},
-l} \|}{\| v^{l,+}_{i,-l-\frac{1}{2}}\|}$ converges to a strictly positive
 limit as $l \raro \infty$ (in fact the above quantity is
 independent of $i$). 

 

  Thus, $|2l+\frac{1}{2}||b^-_{l+\frac{1}{2}}(l,-l-\frac{1}{2})|
\frac{\| v^{l+\frac{1}{2},-}_{i+\frac{1}{2},
 -l} \|}{\|
v^{l,+}_{i,-l-\frac{1}{2}}\|}$ is clearly unbounded as $l$ grows,
  proving
our claim that $[D,\tilde{t}^{\frac{1}{2}}_
 {\frac{1}{2},\frac{1}{2}}]$ is
unbounded.
 \ermrk

\brmrk
  We have verified that $(\cla^\infty, \clh, |D|)$ is a spectral triple
 in the sense of Connes. In fact, this is a ``compact" one, since by construction
 $D$ has compact resolvents. But such a spectral triple with a positive Dirac operator
  is very bad from an algebraic or topological point of view, since the
Fredholm module associated with the sign of $|D|$ is trivial, and hence its
pairing with the $K$-theory is trivial too. Thus, in some sense this spectral
triple fails to capture ``topological" information of the underlying
noncommutative space, i.e. the $C^*$-algebra.
  \ermrk

     However, if we forget for
the time being the topological aspects and concentrate on
      the more
geometric aspects such as the volume form, then it is $|D|$ and not $D$  itself 
which is important. But even in this respect, the present example has
subtle differences from Connes'
  standard formulation, and we shall study
these issues now, leading to a modification of
  Connes' definition and
methods. First of all, let us note that if we take the usual prescription of
noncommutative
  geometry to obtain the ``volume form", i.e. if we take the
functional
  $a \mapsto Lim_{t \raro 0+} \frac{Tr(a exp(-tD^2))}{exp(-tD^2)}$
(where ``Lim"
  is a suitable Banach limit discussed in \cite{F1}), then we
shall get a trace by the general theory, and
  hence the above functional
cannot be the canonical Haar state $\psi$. So, the
  natural question is : how
to recover $\psi$ from the above spectral data ?
   The answer which we are
going to provide requires an additional information, namely
  an operator
$\rho$ described in the following fundamental theorem.
\bthm
 
\label{1}
 Let $B$ be any bounded operator on $h$ such that
$B\tilde{t^n_{ij}}=\lambda_n
  \tilde{t^n_{ij}} \forall n,i,j$, and let $\rho$
denote the operator on $h$
  given by $\rho(\tilde{t^n_{ij}})=q^{-2i-2j}
\tilde{t^n_{ij}}$. Assume that
  $\rho B$ is trace-class (i.e. it has a
bounded extension which is trace-class), and
   define
  a functional $\phi$
on $\cla$ by $\phi(a)=Tr(a\rho B)$. then we have that
  $\phi(a)=\psi(a)
\phi(1)$.
 
\ethm
{\it Proof :-}\\
We first recall the results obtained by  Baaj and Skandalis \cite{BS}, from which it
 follows that there is a unitary operator $W$ acting on $h \ot h$, such that
 $\Phi(a)=W(a \ot I_h)W^*$. Furthermore, it can also be verified that
 $W^*(c \ot 1)=(id \ot \kappa)\Phi(c)$ for any $c \in \cla^\infty$, viewed as
 an element of $h$. Thus, in particular, $W^*(t^n_{ij}\ot 1)=\sum_k t^n_{ik}
 \ot (t^n_{jk})^*$. Now, using the notation of \cite{W1}, we denote by
 $\phi * \psi$ the functional $a \mapsto (\phi \ot \psi)(\Phi(a))$. It
 follows from the definition of the Haar state that
 $(\phi * \psi)(a)=\psi(a)
 \phi(1).$ But on the other hand,
 \bean
 \lefteqn{(\phi * \psi)(a)}\\
 &=&\sum_{n,i,j}
 <\tilde{t^n_{ij}} \ot 1, W(a \ot 1)W^*((\rho B \tilde{t^n_{ij}}) \ot 1)>\\
&=&\sum_{n,i,j,k,l} [2n+1]_q q^{-2i}<t^n_{ik} \ot (t^n_{jk})^*, q^{-2i-2j}
 \lambda_n (at^n_{il}) \ot (t^n_{jl})^*>\\
 &=& \sum_{n,i,j,k}<\tilde{t^n_{ik}}, a
 \tilde{t^n_{ik}}> q^{-2i-2j-2k}\lambda_n [2n+1]_q^{-1}\\
 &=&
 \sum_{n,i,k} <\tilde{t^n_{ik}}, a \rho B(\tilde{t^n_{ik}})>([2n+1]_q^{-1}
 \sum_j q^{-2j}  )=Tr(a\rho B)=\phi(a).\\
 \eean
  Here we have used the facts that
 $ <(t^n_{jk})^*,(t^n_{jl})^*>=\delta_{kl}q^{-2k}[2n+1]^{-1}_q$, and
 $\sum_{j=-n,-n+1,...,n} q^{-2j} =[2n+1]_q.$
\fbox \\

Thus, we obtain the Haar state, i.e. the canonical volume form in this case,
 as follows :\\
\bthm
\label{haar}
 Let $R$ be the operator on $\clh$ defined by $R=I_2 \ot \rho$.
Then we have the following \\
(1) $R$ is a positive unbounded operator having an unbounded inverse $R^{-1}$,
  such that $\cld:=C^2 \ot_{alg} \cla^\infty \subseteq \clh$ is an invariant core
   for $R$,\\
 (2) $\cld$ is also an invariant core for $|D|$ and on this domain
$R$ commutes with $|D|$,\\
(3) For $a \in \cla^\infty$,
  there is a  unique bounded extension of $RaR^{-1}$ belonging to $\cla^\infty$,\\
  (4)    For any $t>0$ and $a \in \cla$,
$$ \frac{Tr(a R exp(-tD^2))}{Tr(R exp(-tD^2))}=\psi(a).$$
Hence in particular, $Lim_{t \raro 0+} \frac{Tr(a R exp(-tD^2))}{Tr(R
 exp(-tD^2))}=\psi(a).$
\ethm
 The proof is straightforward and hence omitted.\\

  We can define an automorphism of $\cla^\infty$ (not adjoint-preserving) given by,
 $\Psi(a)=R a R^{-1}$. It can be shown that $\psi(ab)=\psi(b\Psi(a)) \forall
 a,b \in \cla^\infty.$ This modular property of the Haar state is well-known, and has been
 studied deeply by Woronowicz.

\brmrk
\label{est}

  Since
 $q >1$,   $[x]_q^2=(q^{2x}+q^{-2x}-2)(q-q^{-1})
^{-2}$ is increasing for $x$ in the positive real line, and thus   we can
estimate $Tr(R exp(-tD^2))= \sum_{m=1,2,3,...} [m]_q^2
e^{-t(\frac{m+1}{2})^2}$ by,
 $$ 2 \int_1^\infty [x-1]_q^2
e^{-t(\frac{x+1}{2})^2}dx
  < Tr(Rexp(-tD^2)) <
 2 \int_0^\infty [x+1]_q^2 e^{-t(\frac{x+1}{2})^2} dx.$$
From this, it follows by a direct and simple calculation that there are
positive constants $C_1,C_2,C_3,k$ depending only on $q$   such that $
t^{-\frac{1}{2}}   C_1(exp(\frac{k}{t})-C_3) 
 < Tr(Rexp(-tD^2)) < C_2  t^{-\frac{1}{2}} 
 exp(\frac{k}{t})$ for sufficiently small $t>0.$

This estimate shows that $Lim_{t \raro 0+}
 t^{\frac{1}{2}}e^{-\frac{k}{t}} Tr(Rexp(-tD^2))$ is nonzero and finite. This
 is not something expected from classical intuition, since for any classical
 $d$-dimensional compact manifold one gets a growth rate of the order $t^{-d/2}$
 of $Tr(e^{-tL})$, where $L$ denotes the square of the Dirac operator.
\ermrk

       It is  shown in \cite{CL} and \cite{CD} that           
 the  approach with ``true" Dirac operator works nicely for $q$ of modulus
$1$.  However, as we have discussed in the present letter, there are problems
to extend a similar construction to the case of positive $q$. On the other
hand, the well-developed operator algebraic theory of compact quantum group
due to Woronowicz and others accommodates the case of positive $q$ and not the
complex ones.   We would like to
mention in this context that very recently (after a previous version of this
article was submitted) Chakraborty and Pal (\cite{CP}) have been able to
construct  spectral triples for $SU_q(2)$ with positive $q$, which have a
nontrivial pairing with the $K$-thoery. However, their construction does not
seem to be an extension of the construction of ``true" Dirac operators for
complex $q$ of modulus $1$, although the absolute value $|D|$ (but not
$sign(D)$) of some of the  Dirac operators constructed by them will be
essentially same as the absolute value of the ``true" Dirac operator
considered by us in the present article.\\

{\bf Acknowledgements}\\
The author    is very much grateful to Prof. K. B. Sinha for many valuable suggestions,
  comments and encouragement. He would also like to express his deep gratitude to Prof. Alain Connes
  for pointing out the need of considering the true Dirac operator and
 kindly clarifying the approach involving the true Dirac operator, together with
   many other important comments  which have helped the author to gain deeper
    insight of the problem. Special thanks are to be given to Dr. A. K. Pal for pointing out
     some mistakes in the early manuscript which led to substantial revision and improvement, and to
      P. S. Chakraborty for numerous suggestions and comments, specially related to Fredholm
       modules and cyclic cohomology.    Finally,
 the author would like to thank the Av. Humboldt Foundation
  for the financial support in the form of a Research Fellowship and  Prof. S. Albeverio
 of I.A.M. (Bonn) for his kind hospitality.

\end{document}